\documentclass{revtex4}

\usepackage{axodraw}
\usepackage{amsmath}
\usepackage{amssymb, amsthm}
\usepackage{epsfig}
\usepackage{subfigure}
\usepackage{slashed}
\usepackage{dsfont} 
\usepackage{pstricks}
\usepackage{color}

\newcommand{\beq}{\begin{equation}}
\newcommand{\eeq}{\end{equation}}
\newcommand{\bdm}{\begin{displaymath}}
\newcommand{\edm}{\end{displaymath}}
\newcommand{\beqa}{\begin{eqnarray}}
\newcommand{\eeqa}{\end{eqnarray}}
\renewcommand{\theequation}{\thesection.\arabic{equation}}

\def\st{\sin\theta}
\def\ct{\cos\theta}
\def\sp{\sin\phi}
\def\cp{\cos\phi}

\def\stsq{\sin^2\theta}
\def\ctsq{\cos^2\theta}

\def\cpsq{\cos^2\phi}
\def\casq{\cos^2\alpha}

\begin{document}
\title{Hypercharge-Universal Topcolor}
\author{Felix Braam}
\email[email: ]{felix.braam@physik.uni-freiburg.de}
\author{Michael Flossdorf}
\email[email: ]{michael.flossdorf@physik.uni-freiburg.de}
\affiliation{Institute of Physics, University of Freiburg, 79104 Freiburg, Germany}
\author{R. Sekhar Chivukula}
\email[email: ]{sekhar@msu.edu}
\author{Stefano Di Chiara}
\email[email:]{dichiara@msu.edu}
\author{Elizabeth H. Simmons}
\email[email: ]{esimmons@msu.edu}
\affiliation{Department of Physics and Astronomy\\ Michigan State University\\ East Lansing, MI 48824, USA}

\date{\today}

\begin{abstract}
We propose and discuss the phenomenology of a Topcolor-assisted Technicolor (TC2) model with a flavor-universal hypercharge sector.
After discussing the symmetry breaking pattern and low-energy effective Lagrangian, we examine various experimental and theoretical constraints, finding that precision electroweak measurements yield the strongest bounds on the model.  We perform  a combined fit to all available $Z$-pole and LEP2 data and find that the goodness of fit for hypercharge-universal topcolor is comparable to that of the Standard Model.  In contrast, TC2 models with a flavor non-universal hypercharge sectors are markedly disfavored by the data.
\end{abstract}

\hfill MSUHEP-071107, FR-THEP-07-11

\maketitle

\section{Introduction}

Although the Standard Model is an excellent low-energy effective theory, that describes data at currently accessible energies extremely well, underlying theoretical problems presented by fundamental scalar bosons, such as the hierarchy and triviality problems, suggest that additional physics must describe the dynamics of electroweak symmetry breaking.  The large mass of the top quark has suggested that formation of a top quark condensate could be responsible for all 
\cite{Nambu:1988mr}\cite{Miransky:1988xi}\cite{Miransky:1989ds}\cite{Marciano:1989mj}\cite{Bardeen:1989ds}\cite{Marciano:1989xd}\cite{Dobrescu:1997nm}\cite{Chivukula:1998wd} or at least part  \cite{Hill:1991at} of electroweak symmetry breaking (for a review, see \cite{Hill:2002ap}).  An interesting class of models involving a role for the top quark in dynamical electroweak symmetry breaking is known as topcolor-assisted technicolor (TC2)  \cite{Hill:1991at,Hill:1994hp}.  In these theories, a technicolor condensate provides masses to the weak vector bosons and an extended technicolor (ETC) sector gives mass to the light quarks and leptons -- and a bottom-quark-sized mass to the top.  The majority of the top quark mass (and a small portion of the mass of the $W$ and $Z$) is due to formation of a top-quark condensate through the dynamics of an extended color gauge sector.  The presence of an extended hypercharge sector ensures that the bottom quark (and other standard fermions) do not also condense. In this paper, we will introduce a new model of this kind, discuss its phenomenology, and compare it to the existing TC2 models.

The gauge group of the TC2 models discussed here is
\begin{equation}
SU(N)_{TC} \otimes SU(3)_{1}\otimes SU(3)_{2} \otimes SU(2)_{W} \otimes U(1)_{1} \otimes U(1)_{2}.
\end{equation}

\vspace{-.25cm}
\hspace{4.25cm} $g_{TC}$ \hspace{1cm} $h_1$ \hspace{1cm} $h_2$ \hspace{1cm} $g$ \hspace{1.2cm} $g'_1$ \hspace{.8cm}  $g'_2$ 

\vspace{.25cm}
\noindent where the name of the gauge charge is given below each group.

The technicolor sector is primarily responsible for electroweak symmetry breaking; technicolor is assumed to reside within an ETC model that is entirely responsible for the masses and mixings of the first and second generation quarks and all of the leptons.  
The technicolor sector is primarily responsible for electroweak symmetry
breaking; as is typical for TC2 models \cite{Hill:2002ap}, we assume a walking technicolor
sector that does not generate large precision electroweak corrections. 
In turn, technicolor is assumed to reside within an ETC model that is 
entirely responsible for the masses and mixings of the first and second 
generations and all of the leptons; because ETC need not generate
large quark masses, the usual FCNC constraints on the ETC scale are
alleviated \cite{Hill:2002ap}.
The third generation quarks can, as discussed shortly, receive additional mass contributions from the topcolor sector.  The strong sector consists of the semi-simple group $SU\!\left(3\right)_1\otimes SU\!\left(3\right)_2$ which is spontaneously broken to its diagonal subgroup (identified with $SU(3)_{QCD}$) at a scale $u$ of order a few TeV.  The third generation quarks always transform under the stronger $SU\!\left(3\right)_1$ group; the color group for the first and second-generation quarks is model-dependent.
The weak interactions are represented by an $SU(2)_W$ group, under which all of the quarks and leptons have their standard charges.  Finally, the hypercharge sector consists of the semi-simple group $U\!\left(1\right)_1\otimes U\!\left(1\right)_2$ which is spontaneously broken to $U\!\left(1\right)_Y$ also at the scale $u$.  Again, the third-generation fermions always transform under $U(1)_1$ while the hypercharge group for the first and second generation fermions is model-dependent.

The symmetry-breaking sequence for the models is thus
\begin{eqnarray}
SU(N)_{TC} \,\, \otimes \,\, SU(3)_1 \,\, \otimes\,\, SU(3)_2 \!\!\!\!\! &\otimes& \!\!\!\!\! SU(2)_{L}\,\, \otimes\,\, {U(1)_1}  \,\, \otimes  \,\, {U(1)_2} \nonumber \\
&\big\downarrow& \  \   u \nonumber \\
SU(N)_{TC} \,\, \otimes \,\, SU(3)_C \!\!\!\!\! &\otimes & \!\!\!\!\! SU(2)_{L}\,\, \otimes \,\,  {U(1)_Y}\\
&\big\downarrow&\ \   v \nonumber \\
SU(N)_{TC} \,\, \otimes &SU(3)_C &\,\, \otimes \,\,U(1)_{\rm em}.\nonumber 
\end{eqnarray}
The spontaneous breaking of color and hypercharge to their diagonal subgroups at the scale $u$ is driven by a condensate that is charged under the strong and hypercharge groups (but not under the weak or technicolor groups).  The breaking of the remaining electroweak group to its electromagnetic subgroup is driven by a combination of a top-quark condensate and a technifermion condensate.  Since there are six gauge groups and three symmetry-breakings (color, hypercharge, electroweak), we expect that 9 parameters will be needed to describe the gauge sector; four are the conventional $\alpha_s$, $\alpha$, $G_F$, and $M_Z$ of the Standard model, while the rest describe new physics.

The formation of the top condensate occurs at an energy below the scale of spontaneous symmetry breaking of the topcolor group, $u$.  Hence, the interactions among top quarks can be described using the Nambu--Jona-Lasinio (NJL) approximation \cite{Nambu:1961fr}\cite{Nambu:1961tp}, in which topgluon exchange is modeled as a four-fermi interaction scaled by the size of the topgluon coupling and mass.   In this approximation, we will be able to write down a Dyson-Schwinger ``gap" equation describing the generation of a dynamical mass for the top quark and to determine the range of gauge couplings for which only the top quark obtains a mass.  We will see that the larger the topgluon mass is compared to the top quark mass, the closer the topcolor coupling will be driven to its critical value, the minimum value for which condensation formation is possible (neglecting the impact of the extended hypercharge group).  This is a form of fine tuning \cite{Chivukula:1996cc}.  In addition, in the NJL approximation the  Pagels-Stokar formula \cite{Pagels:1979hd} relates  the scale $\Lambda$ where an interaction becomes strong (approximately the Topcolor scale), the (top-)pion decay constant $f$, and the dynamical mass $m$ (in our case the top mass) generated by the formation of the condensate:
\beq
f^2=\frac{3 m^2}{8 \pi^2} \ln{\frac{\Lambda^2}{m^2}},
\eeq
in the limit $m \ll \Lambda$.   Because the top quark's mass is known, this relationship will allow us to determine the size of the top condensate's contribution to electroweak symmetry breaking.  We will find that electroweak symmetry breaking is still driven primarily by technicolor as the name TC2 suggests.  Effects of the uncertainty in the Pagels-Stokar estimate of $f_t$ are discussed in section \ref{sss:effect}

In the original or ``classic" TC2 model  \cite{Hill:1991at,Hill:1994hp}, both the color and hypercharge interactions are flavor non-universal, treating the third-generation fermions differently than the lighter fermions as shown in Table \ref{assignments-old} (left).   The lack of flavor-universality leads to strong constraints from flavor-changing neutral currents (FCNC) and b-quark physics as discussed in \cite{Buchalla:1995dp}\cite{Burdman:1997pf}\cite{Burdman:2000in}.  In ``flavor-universal" TC2 \cite{Chivukula:1996yr}\cite{Lane:1998qi}\cite{Popovic:1998vb}, which has a flavor-universal color sector but flavor non-universal hypercharge, as in Table \ref{assignments-old} (right), the flavor constraints are less stringent \cite{Popovic:1998vb}\cite{Simmons:2001va}.  
\begin{table}[ht] 
\centering
\begin{tabular}{c|cccccc} 
\multicolumn{6}{c}{\bf Classic Topcolor}\\
\hline \hline \vspace{-0.30cm}\\
      &  ${\mathbf{SU(3)_1}}$ &  ${\mathbf{SU(3)_2}}$ & $SU(2)_W$ & ${\mathbf{U(1)_1}}$ & ${\mathbf{U(1)_2}}$ \vspace{0.05cm} \\ \hline
   \vspace{-0.2cm}\\
    I & - &{\bf SM} & SM & -  & {\bf SM} \\
    II & - & {\bf SM} & SM & -& {\bf SM}  \\    
    III & {\bf SM} & -  & SM & {\bf SM} & - \\ \hline
    \end{tabular}
    \hspace{1cm}
    \begin{tabular}{c|cccccc} 
 \multicolumn{6}{c}{\bf Flavor-Universal Topcolor}\\
\hline \hline \vspace{-0.30cm}\\
      &  ${\mathbf{SU(3)_1}}$ &  ${\mathbf{SU(3)_2}}$ & $SU(2)_W$ & ${\mathbf{U(1)_1}}$ & ${\mathbf{U(1)_2}}$ \vspace{0.05cm} \\ \hline
   \vspace{-0.2cm}\\
     I & {\bf SM}& - & SM & -  & {\bf SM} \\
    II &{\bf SM} & - & SM & -& {\bf SM}   \\    
    III & {\bf SM} & - & SM & {\bf SM} & - \\ \hline  
\end{tabular}
\caption{Gauge charge assignments for fermions of generations I, II and III in the classic \protect\cite{Hill:1991at,Hill:1994hp} and flavor-universal \protect\cite{Popovic:1998vb} Topcolor models.  The entry ``SM'' indicates a charge assignment corresponding to that in the Standard Model. A dash indicates that the fermion is not charged under the gauge group.}
\label{assignments-old}    
\end{table} 
%

Since those models were introduced, a wealth of experimental data on precision electroweak observables has been accumulated -- which now provides the strongest source of constraints on TC2.  Accordingly, 
this paper introduces a third TC2 model, termed ``hypercharge-universal'', in which the extended color interactions single out the third generation (as in classic TC2) but the extended hypercharge interactions are generation universal, as shown in Table \ref{assignments}. We will demonstrate that this model is in better accord with precision electroweak data than the previous TC2 models that featured generation non-universal $Z'$ bosons.
%
\begin{table}[ht] 
\centering
\begin{tabular}{c|cccccc} 
\multicolumn{6}{c}{\bf Hypercharge-Universal Topcolor}\\
\hline \hline \vspace{-0.30cm}\\
      &  ${\mathbf{SU(3)_1}}$ &  ${\mathbf{SU(3)_2}}$ & $SU(2)_W$ & ${\mathbf{U(1)_1}}$ & ${\mathbf{U(1)_2}}$ \vspace{0.05cm} \\ \hline
   \vspace{-0.2cm}\\
    I & - & {\bf SM} & SM & {\bf SM} & - \\
    II & - & {\bf SM} & SM & {\bf SM} & - \\    
    III & {\bf SM} & - & SM & {\bf SM} & - \\ \hline
\end{tabular}
\caption{Gauge charge assignments for fermions of generations I, II and III in the hypercharge-universal topcolor model introduced in this paper.  The entry ``SM'' indicates a charge assignment corresponding to that in the Standard Model. }
\label{assignments}    
\end{table} 
%

In the next section, we analyze topcolor and hypercharge symmetry breaking in the hypercharge-universal topcolor-assisted technicolor model, including a brief discussion of constraints on the effective theory below the topgluon mass. In section III  we analyze electroweak symmetry breaking in the model, and in section IV we
present  an analysis of the constraints arising from precision electroweak data.
Section V updates the constraints on the TC2 models with flavor non-universal $Z'$ bosons.  Our conclusions are presented in section VI.

\section{The Hypercharge-Universal Topcolor Model}

We now discuss the spontaneous breaking of the extended color and hypercharge groups and construct the effective Lagrangian describing physics below the topgluon mass scale.  We show that, in the NJL approximation, the model (like classic and flavor-universal TC2) does admit the possibility of forming a top condensate without having other fermion condense.  We also estimate the constraint on the model's parameter space from the need to avoid the $U(1)_1$ Landau pole, remain consistent with LEP2 limits on four-fermion contact interactions, and avoid large FCNC.

\subsection{Topcolor Symmetry Breaking} \label{Sec:BreakingSymmetry}

At the scale $u$, $SU(3)_{1} \otimes SU(3)_{2}$ and $U(1)_{1} \otimes U(1)_{2}$ break to their respective diagonal subgroups, which we associate with $SU(3)_{QCD}$ and $U(1)_Y$.  The condensate
responsible for this symmetry breaking is taken to transform as a 
$\left(3,\bar{3},\frac{p}{2\sqrt{6}},-\frac{p}{2\sqrt{6}}\right)$ under these interactions,
where $p$ is an arbitrary $U(1)$ charge of order 1.
The mass-squared matrix for the colored gauge-bosons is then
\begin{equation}
\frac{u^2}{2}   
\left(
 \begin{array}{cc}
h_1^2&-h_1h_2\\
-h_1h_2&h_2^2\\
\end{array}
\right)~.
\label{MassMatrix}
\end{equation}
Defining $\cos{\eta} \equiv h_1/\sqrt{h_1^2 + h_2^2}$ and $\sin{\eta} \equiv h_2/\sqrt{h_1^2 + h_2^2}$, we obtain for the mass eigenstates $C_\mu^A$ (massive topgluons) and $G_\mu^A$ (massless fields identified with the QCD gluons), the relations
\begin{equation}
C^A_{\mu} = \cos{\eta} \, A^A_{1 \mu} - \sin{\eta} \, A^A_{2 \mu}
\qquad\qquad G^A_{\mu} = \sin{\eta} \, A^A_{1 \mu} + \cos{\eta} \, A^A_{2 \mu}.
\label{MassEigenstates}
\end{equation}
The mass of the topgluon is
\begin{equation}
M_C \,=\,u \sqrt{h_1^2 + h_2^2} ~,
\label{primeMassesC}
\end{equation}
and the coupling of the gluon ($g_S$) is given by
\begin{equation}
g_S \equiv \frac{h_1 h_2}{\sqrt{h_1^2+h_2^2}} = h_1 \sin\eta = h_2\cos\eta~.
\label{eq:gcouplS}
\end{equation}

In the limit $v \ll u$,  a similar analysis can be made for the $U(1)$ fields to zeroth order in $v/u$.
The $2 \times 2$ mass-squared matrix for the $U(1)$ fields is given by
\begin{equation}
\frac{p^2\,u^2}{2}   
\left(
 \begin{array}{cc}
{g'}_1^2&-{g'}_1{g'}_2\\
-{g'}_1{g'}_2&{g'}_2^2\\
\end{array}
\right)~.
\label{MassMatrixp}
\end{equation}
Defining $\cos \phi = {g'}_1/\sqrt{{g'}^2_1 + {g'}^2_2}$ and $\sin\phi = {g'}_2/\sqrt{{g'}^2_1 + {g'}^2_2}$, we
may define the $U(1)$ fields
\begin{equation}
Z'_\bot = \cos{\phi}\ B_2^\mu + \sin{\phi}\ B_1^\mu \quad\quad
Z' = -\sin{\phi}\ B_2^\mu + \cos{\phi}\ B_1^\mu~.
\label{MassEigenstatesZ}
\end{equation}
Before electroweak symmetry breaking occurs, the $Z'_\bot$ corresponds to the (massless) 
standard model hypercharge field, which couples to $Y=Y_1+Y_2$, with strength 
\begin{equation}
g' \equiv \frac{g'_1 g'_2}{\sqrt{g_1^{\prime 2}+g_2^{\prime 2}}}= g'_1 \sin\phi  = g'_2\cos\phi~.\nonumber
\label{eq:gcouplY}
\end{equation}
The orthogonal field, $Z'$, acquires the mass
\begin{equation}
M_{Z'} \,=  \frac{p\, u}{2} \sqrt{g_1^{\prime 2} + g_2^{\prime 2}}~.
\label{primeMassesZ}
\end{equation}
A complete analysis of the neutral gauge-boson sector, including electroweak
symmetry breaking, will be considered below in section \ref{EWSymmetryBreaking}.

\subsection{The Topgluon and $Z'$ Effective Lagrangians}\label{effLag}

At energies below scale $u$, exchange of the massive topgluon and $Z'$ bosons between fermion currents can be described by a low-energy effective Lagrangian.   We will derive the form of this effective Lagrangian for the Hypercharge-Universal Topcolor model; four-fermion interactions induced by topgluon exchange will treat the third generation differently while those induced by $Z'$ exchange will be flavor-universal in form.  As described in \cite{Hill:1994hp}\cite{Buchalla:1995dp}\cite{Popovic:1998vb}, the effective Lagrangians for the classic and flavor-universal Topcolor models differ from the one we derive here, due to the alternative fermion charge assignments.

The fermion kinetic energy term includes interactions between fermion currents and the mass-eigenstate topgluon and gluon fields
\begin{equation}
\sum_{i=I,II,III} \bar{q}_{i} i \slashed{D} q_{i}  = \ldots - C_\mu^A J^{A \mu}_C - G_\mu^A J^{A\mu}_G,
\end{equation}
where $q_i$ denotes a quark in generation $i$ and the currents are defined as follows, with the fermion charges of table \ref{assignments} taken into account: 
\begin{eqnarray}
&J^{A \mu}_C \equiv g_S \cot{\eta} \,\bar{q}_{III} \gamma^\mu \frac{\lambda^A}{2}q_{III} - g_S \tan{\eta}\, (\bar{q}_{II} \gamma^\mu \frac{\lambda^A}{2}q_{II} + \bar{q}_{I} \gamma^\mu \frac{\lambda^A}{2}q_{I}) \label{currentC} \\
&J^{A\mu}_G \equiv g_S \sum_i \left( \bar{q}_{i} \gamma^\mu \frac{\lambda^A}{2}q_{i} \right) \label{currentG}.
\end{eqnarray}
If we define 
\begin{equation}
 \kappa_3 \equiv \alpha_S \left(\frac{h_1}{h_2}\right)^2 = \alpha_S \cot^2\eta
\end{equation}
with $\alpha_S=g^2_S/4\pi$, 
then eqn.\ (\ref{currentC})  can be rewritten
\begin{equation}
J^{A \mu}_C = \sqrt{4 \pi} \left(\sqrt{\kappa_3} \bar{q}_{III} \gamma^\mu \frac{\lambda^A}{2}q_{III}-\frac{\alpha_S}{\sqrt{\kappa_3}} (\bar{q}_{II} \gamma^\mu \frac{\lambda^A}{2}q_{II}+\bar{q}_{I} \gamma^\mu \frac{\lambda^A}{2}q_{I})\right)\,.
\end{equation}
At energies well below the topgluon mass,  one obtains the effective Lagrangian term 
\begin{equation}
\mathcal{L}_C = - \frac{1}{2 M_C^2} J_{C \mu}^{A} J^{A \mu}_C
\label{effLag1}
\end{equation}
which includes the symmetry factor for two identical currents. 
The corresponding result for the hypercharge sector is
\begin{equation}
\mathcal{L}_{Z'} =-\frac{2 \pi}{M_{Z'}^2}\kappa_1 \left[ \sum_{i=I,II,III} \left(\bar{f_i}\gamma^\mu Y f_i \right)\right]^2,
\label{effLag2}
\end{equation}
\begin{equation}
\kappa_1 \equiv \alpha_Y \left(\frac{g'_1}{g'_2}\right)^2\,.
\end{equation}
where $\alpha_Y=g'^2/4\pi$ and $f_i$ denotes any quark or lepton of generation $i$.

We require four parameters, in addition to those of the standard model, to describe the new color and hypercharge physics discussed so far (two extra gauge couplings, the new symmetry-breaking scale, and the hypercharge of the condensate associated with that scale).  We choose to write these as  $\kappa_1, \kappa_3, p$ and the scale $u$ to be consistent with \cite{Popovic:1998vb}. We therefore re-express the topgluon  (\ref{primeMassesC}) and the $Z'$ masses (\ref{primeMassesZ}) in terms of these parameters, using Eqns. (\ref{eq:gcouplS}) and (\ref{eq:gcouplY}). 
The following equalities
\begin{equation}
\begin{split}
h_1^2 &= 4 \pi \left(\kappa_3+\alpha_S\right),\qquad\qquad
h_2^2 = 4 \pi \frac{\alpha_S}{\kappa_3} \left(\kappa_3+\alpha_S\right)\\
{g'_1}^2 &= 4 \pi \left(\kappa_1+\alpha_Y\right),\label{g1'} \qquad\qquad \!\!\! {g'_2}^2 = 4 \pi \frac{\alpha_Y}{\kappa_1} \left(\kappa_1+\alpha_Y\right),
\end{split}
\end{equation}
allow us to write 
\beq
\begin{split}
M_{C}&= u\, \sqrt{\frac{4 \pi}{\kappa_3}}\left(\kappa_3+\alpha_S\right),\\
M_{Z'}&=p\, u\sqrt{\frac{\pi}{\kappa_1}}\left(\kappa_1+\alpha_Y\right).\label{kappaMasses}
\end{split}
\eeq

\subsection{The Gap Triangle}\label{GapEquations}

We now introduce the gap equation that describes the self-consistent dynamical generation of fermion masses in the Nambu--Jona-Lasinio  approximation \cite{Nambu:1961fr}\cite{Nambu:1961tp}.  This has the overall form of the gap equation in  ref.\ \cite{Popovic:1998vb}, but the coupling factors are modified to reflect the universal $Z'$ and non-universal topgluons of the hypercharge-universal model:
\begin{equation}
m_f = G_1 \frac{m_f M_{Z'}^2}{8 \pi^2}\left[1-\left(\frac{m_f}{M_{Z'}}\right)^2 \ln{\left(\frac{M^2_{Z'}}{m_f^2}\right)}\right] +\, G_3 \frac{3 m_f M_{C}^2}{8 \pi^2}\left[1-\left(\frac{m_f}{M_{C}}\right)^2 \ln{\left(\frac{M^2_{C}}{m_f^2}\right)}\right],
\label{mf}
\end{equation}
where the coefficients $G_i$ are
\bdm
\begin{array}{llll}
G_1&=&\frac{8\pi}{M^{2}_{Z'}}\kappa_1Y^{f}_{L}Y^{f}_{R}&\mbox{for all fermions,}\\
G_3&=&0 &\mbox{for leptons}, \\
&=&\frac{4\pi\alpha^{2}_{S}}{M^{2}_{C}}\frac{1}{\kappa_3} & \mbox{for quark generations I and II},\\
&=&\frac{4\pi}{M^{2}_{C}}\kappa_3 & \mbox{for quark generation III}.\label{Gi}
\end{array}
\edm
and we define hypercharge as $Q = T_3 + Y$. In order for a fermion to condense, its gap equation (\ref{mf}) must have a solution for $m_f \!\!>\!\! 0$. Accordingly we expect condensation, only if
\beq
1 < \frac{1}{8 \pi^2} \left(G_1 M_{Z'}^2 + 3 G_3 M_C^2\right).
\label{eq:Gi}
\eeq
Evaluating this for each fermion species yields constraints on $\kappa_1$ and $\kappa_3$.  The three most stringent are: 
\begin{align}
\kappa_3 + \frac{2}{27}\kappa_1 &> \frac{2}{3} \pi, \label{c1}\\
\kappa_3 - \frac{1}{27} \kappa_1 &< \frac{2}{3} \pi, \label{c2}\\
\kappa_1 &< 2 \pi \label{c3}.
\end{align}
Equation (\ref{c1}) must hold if  the top quark is to condense; equations (\ref{c2}) and (\ref{c3}) must be fulfilled in order to avoid  bottom quark or $\tau$-lepton condensation, respectively. In contrast, avoiding charm quark condensation, for instance, imposes the less stringent constraint  
\begin{equation}
\kappa_1 + \frac{27}{2} \frac{\alpha_S^2}{\kappa_3} < 9 \pi.
\label{c4}
\end{equation}

The above gap equations only take into account the $Z'$ and topgluon interactions in the NJL Lagrangian.  A more complete description would also include the electromagnetic and the QCD interaction by using the gauged NJL model \cite{Bardeen:1985sm}\cite{Leung:1985sn}.   The result is a shift  \cite{Appelquist:1988vi}\cite{Yamawaki:1988na}\cite{Kondo:1988qd}\cite{Miransky} of the equations (\ref{c1}-\ref{c3}), analogous to the findings in ref.\ \cite{Popovic:1998vb}:
\begin{align}
\kappa_3 + \frac{2}{27}\kappa_1 &> \frac{2}{3} \pi - \frac{4}{3} \alpha_S - \frac{4}{9} \alpha_Y, \label{c1'}\\
\kappa_3 - \frac{1}{27} \kappa_1 &< \frac{2}{3} \pi - \frac{4}{3} \alpha_S + \frac{2}{9} \alpha_Y,  \label{c2'}\\
\kappa_1 &< 2 \pi - 6 \alpha_Y \label{c3'}.
\end{align}
The difference is not great for larger values of $\kappa_1$ but is important when $\kappa_1$ is small; as we shall shortly see, experiment favors the latter.  

\begin{figure}[ht]
\begin{center} 
\epsfig{file=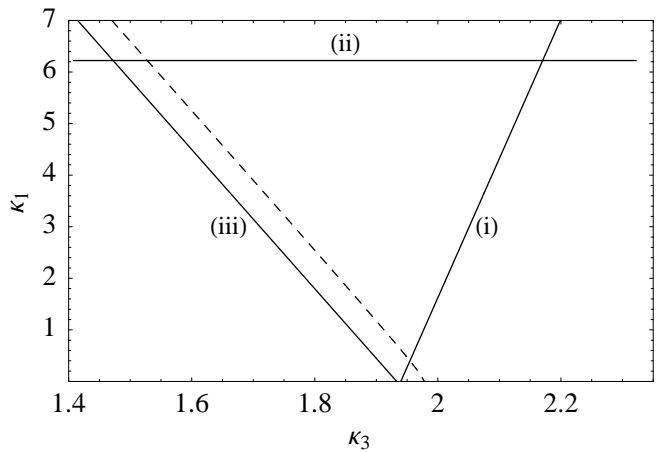,height=6cm}
\caption{\label{GapTriangleGauged}  The Gap Triangle corresponding to eqns.\ (\ref{c1'}-\ref{c3'}) lies between the solid lines (i) the lower bound for $\langle \bar{b}b\rangle = 0$, (ii) the upper bound for $\langle \bar{\tau}\tau\rangle = 0$ and (iii) the lower bound for $\langle \bar{t}t\rangle \neq 0$. The values $\alpha_Y\!\left(M_Z\right) = 0.010$ and $\alpha_S\!\left(M_Z\right) =0.118$ from ref.\ \protect\cite{Yao:2006px} were used in the calculations.  The dashed line represents  a solution of the gap equation for the scale $u=500 \ {\rm GeV}$ and $p=1$ for a dynamical top mass of $m_t\approx170 \ {\rm GeV}$. Solutions for higher scales lie to the left of this line.}
\end{center} 
\end{figure}

The three inequalities are simultaneously satisfied inside a nearly-triangular region (the ``gap triangle") in the $\kappa_1 - \kappa_3$ plane, as shown in Figure \ref{GapTriangleGauged}. 
While the entire gap triangle can formally allow a top condensate to be generated, other considerations indicate that the phenomenologically relevant region lies near the left-hand boundary determined by eqn. (\ref{c1'}), as we shall now discuss.
The dimensionful scales in the gap equation, eqn. (\ref{mf}) are $M_{C,Z'}$, and are hence of order
$u$. For generic values of the couplings which satisfy eqn. (\ref{c1'}), therefore, one will obtain
a top quark mass of order $u$. For our theory to approximate the standard model, however,
 $u$ must lie significantly above the weak scale. To obtain $m_t \ll u$, therefore,
 the model must be in the area adjacent to the left-hand side of the gap triangle.  Using eqns.\ (\ref{kappaMasses}) in the gauged  NJL version of eqn.\ (\ref{mf}) gives a relation between $\kappa_1$ and $\kappa_3$ for a given (desired) dynamical fermion mass $m_f$, scale $u$ and hypercharge factor $p$. Since the top mass is known to be $(172.5\pm 2.3)\ {\rm GeV}$ \cite{Yao:2006px} and the dynamical mass contribution coming from Extended Technicolor can be estimated \cite{Hill:1994hp} to be of the order of a few GeV, the Topcolor sector of our model must give a dynamical mass contribution of around $170 \ {\rm GeV}$. The dashed line in Fig.\ \ref{GapTriangleGauged} shows the solution to the gauged version of the gap equation (\ref{mf}) with this dynamical mass for $p=1$ and  $u=500 \ {\rm GeV}$; solutions for scales above $500 \ {\rm GeV}$ would lie even closer to the left hand side of the gap triangle, which is approached in the limit $u\rightarrow \infty$.

\subsection{Preliminary Constraints}

Next, we will discuss the constraints on the parameters of the hypercharge-universal topcolor-assisted
technicolor model arising from having  sufficiently weakly-coupled $U(1)$ gauge interactions,
from searches for contact-interactions, and from the absence of large flavor-changing
neutral currents.

\subsubsection{Absence of a Landau Pole}

To avoid conflict with the Landau pole for hypercharge group  $U\!\left(1\right)_1$, the value of $\kappa_1$ must be relatively small.  Just as in QED, fermion-loop corrections to the gauge boson propagator cause the coupling constant $g'_1$ of the $U\!\left(1\right)_1$ group \footnote{Since no fermions are charged under $U\!\left(1\right)_2$ there will be no running of the coupling constant $g'_2$.} to run. Working in the $\overline{\rm MS}$ renormalization scheme, in which the 
 running coupling is independent of the fermion masses, the beta function is
\beq
\beta\!\left(\mu\right) \equiv \mu \frac{\partial g'_1}{\partial \mu} = C \ \frac{{g'_1}^3}{12 \pi^2},\label{beta} \qquad\qquad C \equiv \frac{1}{2} \sum_i Y_i^2
\eeq
where $\mu$ is the energy scale in $\overline{\rm MS}$ and $C$ 
is the sum over the squared hypercharges of all fermions (left and right handed).   Summing over the charges of all three fermion generations, we find $C=5$ for the hypercharge-universal topcolor model. 
Integrating eqn.\ (\ref{beta}) yields the scale dependent coupling at an arbitrary scale $\Lambda$, if we know the coupling at another scale $\Lambda_0$:
\beq
{g'_1\!\left(\Lambda\right)}^2= {g'_1\!\left(\Lambda_0\right)}^2 \ \frac{1}{1-\frac{C}{6 \pi^2}\ {g'_1\!\left(\Lambda_0\right)}^2\ln{\frac{\Lambda}{\Lambda_0}}}.\label{Running}
\eeq
The running coupling thus exhibits a Landau pole for $\Lambda=\Lambda_L$,
\beq
\Lambda_L = \Lambda_0 \exp{\left[{\frac{6 \pi^2}{C {g'_1\!\left(\Lambda_0\right)}^2}}\right]} = \Lambda_0 \exp{\left[{\frac{3 \pi}{2C (\kappa_1 (\Lambda_0) + \alpha_Y(\Lambda_0))}}\right]}\,,
\eeq
where eqn. (\ref{g1'}) has been used to obtain the right-most expression. For the theory to be self-consistent, the Landau pole $\Lambda_L$ should be at an energy scale well above the symmetry breaking scale $u$.   Taking $\Lambda_0 \approx u$ and $C=5$, we find that if $\Lambda_L$ is to lie at least 2 orders of magnitude above $u$, we must require $\kappa_1 \lesssim 0.2$.  This constraint will be superseded by the precision electroweak bounds.

\subsubsection{Limits from Four-Fermion Contact Interactions}\label{ContactInt}

Experimental limits on contact interactions provide a lower limit on the $Z'$ mass as a function of  $\kappa_1$.  Published analyses \cite{Yao:2006px} of the relevant data assumes an effective Lagrangian of the general form
\begin{equation}
\frac{4\pi}{(1 + \delta) (\Lambda^\pm_{ij})^2}\sum_{i,j = L,R} \eta_{ij} \left(\bar{e}_i \gamma_\mu e_i\right) \left(\bar{f}_j \gamma^\mu f_j\right) ,
\end{equation}
where $\delta = 1$ if $f$ is an electron and $\delta = 0$ otherwise.  The coefficient $\eta_{ij}$ has the value $\pm 1$ corresponding to constructive (destructive) interference.   Comparing this effective Lagrangian with  the relevant terms of the low-energy interaction due to $Z'$ exchange in our model (\ref{effLag2}), we find the correspondence
\begin{equation}
M_{Z'} = \Lambda^{sgn[Y_{f_i}]}_{ij}  \sqrt{\kappa_1 Y_{e_i} Y_{f_j}}
\label{eq:mzlam}
\end{equation}

The LEP Electroweak Working Group (LEP EWWG) analysis of possible contact interaction contributions to $e^+e^- \to f \bar{f}$ \cite{Alcaraz:2006mx} quotes experimental lower bounds on the various $\Lambda^{sgn[Y_{f_i}]}_{ij} $ corresponding to different fermion chiralities and flavors.  Given the factors of hypercharge appearing in eqn. (\ref{eq:mzlam}), the strongest of the LEP EWWG limits on this model comes from the process $e^+_R e^-_R \to \ell^+_R \ell^-_R$ where $\ell = \mu, \tau$:
\beq
\Lambda^-_{RR} \geq 9.3\, {\rm TeV}\quad {\rm corresponding\ to} \quad M_Z' \geq  9.3\, \sqrt{\kappa_1}\,{\rm TeV}
\eeq
A separate analysis reported in ref. \cite{Cheung:2001wx} combines data from HERA, LEP, and atomic parity violation to reach limits on new contact interactions; in this case, the strongest limit for this model comes from $e^+_R e^-_R \to u_R u_R$:
\beq
\Lambda^+_{RR} \geq 17.9\, {\rm TeV}\quad {\rm corresponding\ to} \quad M_Z' \geq 14.6\, \sqrt{\kappa_1}\,{\rm TeV}~.
\eeq
We will find in Section IV that precision electroweak tests impose even stronger constraints on $M_{Z'}$.

\subsubsection{Flavor Changing Neutral Currents: $K\bar{K}$ and $B\bar{B}$ mixing}\label{sec:FCNC}

The effective Lagrangian for topgluon exchange, eqn. (\ref{effLag1}), induces flavor changing neutral currents (FCNC) at tree level. To estimate the size of the FCNC, we assume a mixing only between the left-handed down-type quark fields (e.g. as consistent with the discussion in \cite{Burdman:2000in}) and use  the CKM matrix of the SM to approximate the size of the mixing.  The relevant piece of (\ref{effLag1}) is then as follows, where the quark fields $d_i$ are mass eigenstates
\begin{equation}
\mathcal{L}=- \frac{2\pi}{M_C^2}   \biggl[  \sqrt{\kappa_3}  \left(  \bar{d}_L^j V^{\dagger j 3} \gamma^\mu \frac{\lambda^A}{2} V^{3 k} d_L^k \right)\biggr.  
-\biggl.\frac{\alpha_S}{\sqrt{\kappa_3}}   \left( \bar{d}^j V^{\dagger j 1}\gamma^{\mu} \frac{\lambda^a}{2} V^{1 k} d_L^k + \bar{d}^j V^{\dagger j 2}\gamma^{\mu} \frac{\lambda^a}{2} V^{2 k} d_L^k\right)\biggr]^2\,.
\end{equation}
The terms contributing to $K^0 \bar{K}^0$ mixing are
\begin{equation}
\mathcal{L}^{Tc}_{\Delta s = 2}  = - \frac{2\pi}{M_C^2}  \biggl[\sqrt{\kappa_3} V^{\dagger 1 3} V^{3 2} - \frac{\alpha_S}{\sqrt{\kappa_3}}\left( V^{\dagger 1 1} V^{1 2} + V^{\dagger 1 2} V^{2 2}\right) \biggr]^2
\times \left(\bar{d}_L\gamma^\mu \frac{\lambda^A}{2} s_L \right)^2\,.
\label{TopColorDbarS}
\end{equation}
Applying a Fiertz transformation and a trace identity for the Gell-Mann matrices yields
\begin{equation}
\mathcal{L}^{Tc}_{\Delta s = 2} = - \Omega^{Tc} (M_C, \kappa_3) \times \left( \bar{d}_{L \alpha} \gamma_\mu s_{L \alpha} \right) \left( \bar{d}_{L \beta} \gamma^\mu s_{L \beta} \right),
\end{equation}
\begin{equation}
\Omega^{Tc} (M_C, \kappa_3)  \equiv \frac{2 \pi}{3 M_C^2} \left[ \sqrt{\kappa_3} V^{\dagger 1 3} V^{3 2} - \frac{\alpha_S}{\sqrt{\kappa_3}}\left( V^{\dagger 1 1} V^{1 2} + V^{\dagger 1 2} V^{2 2}\right)\right]^2\,.
\end{equation}

Rather than using this result to compute the topcolor contribution to the $K^L K^S$ mass difference directly, we will compare the coefficient $\Omega^{Tc}$ to the coefficient $\Omega^{SM}$ that multiplies the same $\Delta S = 2$ operator in the SM:
\begin{equation}
\Omega^{SM} \equiv \frac{G_F}{\sqrt{2}}\frac{\alpha}{4 \pi} \frac{1}{\sin^2{\theta_W}} \left[ \sum_i \left(V^{\dagger 1 i} V^{i 2}\right)^2 x_i + \sum_{i \neq j} V^{\dagger 1 i} V^{i 2} V^{\dagger 1 j} V^{j 2}  \frac{x_i x_j}{x_i - x_j} \ln{\frac{x_i}{x_j}}\right]\,,
\end{equation}
where $x_i\equiv m_i^2 / M_W^2$ and the indices $i, j$ run over the up, charm and top quark flavors.
In order to retain agreement with experiment, the extra contribution from topcolor should not exceed that already present in the SM (i.e. $\Omega^{Tc} \leq \Omega^{SM}$).  We effect the comparison using the following values of the magnitudes of the CKM elements, calculated from the measured \cite{Yao:2006px} $\vert V_{ud}\vert$, $\vert V_{ub} \vert$ and $\vert V_{cb} \vert$ using the standard parametrization of the CKM matrix \cite{Eidelman:2004wy}
\begin{equation}
V \approx \left(\begin{array}{ccc}
      0.9738& 0.2275& 0.00431\\
      -0.2275& 0.9729& 0.0416\\
      0.005271&  -0.04149& 0.9991
\end{array}\right)\,.
\label{CKMmeassured}
\end{equation}

\begin{figure}[hbt]
\begin{center}
\epsfig{file=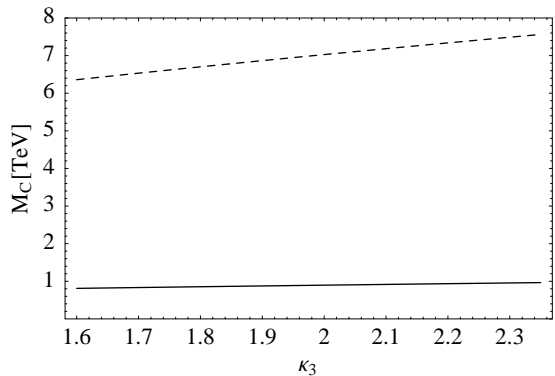,height=5cm} 
\caption{\label{BBbarKKbarMcConstraint}  Lower bound for $M_C$ due to $K\bar{K}$ (solid line) and  $B\bar{B}$ (dashed line) mixing as a function of $\kappa_3$.   }
\end{center} 
\end{figure}

The solid line in Fig.\ \ref{BBbarKKbarMcConstraint} shows the resulting lower bound on $M_C$ as a function of $\kappa_3$, if we require $\Omega^{Tc} \approx  \Omega^{SM}$ and set $\alpha_S\!\left(M_Z\right) = 0.118$. Due to this constraint, $M_C \gtrsim 1 \,{\rm TeV}$.
The analogous calculation for $B \bar{B}$ mixing (making the replacements $s \to d$ and $d \to b$) yields a stronger constraint shown by the dashed line of Fig.\ \ref{BBbarKKbarMcConstraint}: $M_C \gtrsim 6 - 8 \,{\rm TeV}$.  This limit is consistent with the results obtained by \cite{Burdman:2000in} for classic TC2 where both the topgluons and $Z'$ contribute to tree-level FCNC.  
In Section IV, these limits will be superseded by those from precision electroweak data.

\section{Electroweak Symmetry Breaking and Variables}

\subsection{Electroweak Symmetry Breaking}\label{EWSymmetryBreaking}

We now consider the electroweak neutral-boson sector in greater detail.  
The complete pattern of electroweak breaking is:
\begin{eqnarray}
\stackrel{W^{a \mu}_{\phantom1_{\phantom*}}}{SU(2)_{L}} \!\!\!\!\! &\otimes&\!\!\!\!\! \stackrel{B_{1_{\phantom*}}^\mu}{U(1)_1} \otimes \stackrel{B_{2_{\phantom*}}^\mu}{U(1)_2} \nonumber \\
&\big\downarrow \  \   u& \nonumber \\
\stackrel{W^{a \mu}_{\phantom1_{\phantom*}}}{SU(2)_{L}}\!\!\!\!\!& \otimes&\!\!\!\!\! \stackrel{Z'_{\bot_{\phantom*}}}{U(1)_Y}\\
&\big\downarrow\ \    v &\nonumber \\
&\stackrel{A^\mu_{\phantom1}}{U(1)_{em}}&.\nonumber \\
\vspace{-0.25cm} \nonumber
\end{eqnarray}
In order to calculate the effects of this extended
gauge structure on electroweak processes, we begin by constructing a new basis for the neutral electroweak gauge-bosons.
The electroweak covariant derivative may be written in terms of the original gauge eigenstates as 
\beq
\partial^\mu + i g \,T^a \,W_{a}^\mu + ig_1' \,Y_1\,B_1^\mu
+ig_2'\,Y_2 B_2^\mu.
\label{CoDerivativeEW}
\eeq
Since the electric charge operator $Q=T_3+Y = T_3+ Y_1+Y_2$ remains  unbroken, it is useful to rewrite (\theequation) in terms of the massless EM-field $A^\mu$ and two orthogonal neutral fields. 
First we use mixing angle $\phi$ to rotate the hypercharge bosons (see eqn. (\ref{MassEigenstatesZ}))
\begin{equation}
Z'_\bot = \cos{\phi}\ B_2^\mu + \sin{\phi}\ B_1^\mu \quad\quad
Z' = -\sin{\phi}\ B_2^\mu + \cos{\phi}\ B_1^\mu,
\end{equation}
where $Z'_\bot$ couples to $Y=Y_1+Y_2$ and is orthogonal to the massive $Z'$ introduced in section \ref{Sec:BreakingSymmetry}.  Then we use angle~$\theta$ to define the states 
\begin{equation}
A^\mu = \cos{\theta}\ Z'^{\mu}_\bot + \sin{\theta}\ W_3^\mu\quad \quad
Z^\mu = -\sin{\theta}\ Z'^{\mu}_\bot + \cos{\theta}\ W_3^\mu.
\end{equation}
In terms of these two mixing angles, the gauge couplings may be expressed as
\beq
g = \frac{e}{\st}\,,\quad\quad
g_1' = \frac{g^\prime}{\sp} = \frac{e}{\sp\ct}\,,\quad\quad
g_2' = \frac{g^\prime}{\cp} = \frac{e}{\cp\ct}\,,
\eeq
where $e$ is the electromagnetic coupling, given by
\begin{equation}
\frac{1}{e^2} = \frac{1}{g^2}+\frac{1}{g'^2_1}+\frac{1}{g'^2_2}~.
\end{equation}
In this rotated basis, we may rewrite the covariant derivative (\ref{CoDerivativeEW}) as
\beq
D^\mu = \partial^\mu + i g \,T^a \,W_{a}^\mu + i e A^\mu Q + i \frac{e}{\st \ct}Z^\mu \left( T^3 - \sin^2{\theta}\, Q\right) +i \frac{e}{\ct \sp \cp} Z^{\prime \mu} Y',\label{covDerivative}
\eeq
where $Y'\equiv Y_1-Y \sin^2{\phi} $.

Following the procedure outlined in ref. \cite{Chivukula:1996cc}, we may now evaluate the mass-matrices for the gauge-bosons. The $W$ boson mass may be written
\begin{equation}
M^2_W = \frac{e^2}{4 \sin^2\theta} \langle T_1 T_1\rangle~, \label{wMass}
\end{equation}
where $\langle T_1 T_1 \rangle$ denotes the electroweak symmetry breaking condensate
that couples to the charged-boson gauge fields. 
Given the size of the $W$ boson mass, we note that
the numerical value of this condensate is $\langle T_1 T_1 \rangle = v^2
\approx (250\ {\rm GeV})^2$. In a topcolor assisted technicolor model, electroweak symmetry-breaking
rises from both a technifermion and top condensate. We therefore
expect
\begin{align}
\langle T_1 T_1\rangle & = \langle T_1 T_1\rangle_{TC} + \langle T_1 T_1 \rangle_{top} \\
& = F^2_{TC} + f^2_t ~,
\end{align}
where $F_{TC}$ and $f_t$ are the technicolor and topcolor $F$-constants, analogous to
$f_\pi$ in QCD. In analogy with the definition of $\tan\beta$ in two-Higgs-doublet models, we 
define an angle $\alpha$ by
$F_{TC}\equiv v \cos\alpha$ and $f_t \equiv v \sin\alpha$.   This angle $\alpha$ (or, equivalently, $f_t$) is the fifth and final parameter needed to describe all of the physics beyond the SM in our model.
Note that $f_t$ is determined by
the topcolor dynamics through the Pagels-Stokar relation \cite{Pagels:1979hd} as mentioned in the Introduction. 

From the gauge-boson couplings in eqn. (\ref{covDerivative}), we find the neutral-boson mass
matrix in the $Z$ -- $Z'$ basis to be
\begin{equation}
{\mathsf m}^2_Z = \frac{e^2}{4\sin^2\theta \cos^2\theta}
\begin{pmatrix}
\langle T_3 T_3 \rangle & \frac{\sin\theta}{\sin\phi \cos\phi} \langle T_3 Y'\rangle \\
 \frac{\sin\theta}{\sin\phi \cos\phi} \langle T_3 Y'\rangle & \frac{\sin^2\theta}{\sin^2\phi \cos^2\phi} \langle Y' Y' \rangle
 \end{pmatrix}~,
 \end{equation}
where we note that electric charge $Q$ is unbroken, and therefore all terms involving
the condensate coupling to $Q$ must vanish. Furthermore, by custodial symmetry,
$\langle T_1 T_1\rangle = \langle T_2 T_2 \rangle = \langle T_3 T_3 \rangle$, allowing us
to evaluate the upper left-hand element as having the value $v^2$.

Next, consider the off-diagonal elements proportional to $\langle T_3 Y'\rangle$,
\begin{align}
\langle T_3 Y' \rangle & = \langle T_3 (Y_1 - Y \sin^2\phi)\rangle \\
& = \langle T_3 (Y_1-(Q-T_3)\sin^2\phi \rangle \\
& = \langle T_3 Y_1\rangle + \langle T_3 T_3\rangle \sin^2\phi~.
\end{align}
This matrix element receives contributions both from technifermion and top condensates.
To minimize the size of electroweak corrections, we will assume that the
technicolor sector couplings to $U(1)_1$ are isospin symmetric \cite{Lane:1995gw},
in which case $\langle T_3 Y_1 \rangle_{TC}=0$, yielding a technicolor
contribution
\begin{equation}
\langle T_3 Y'\rangle_{TC} = v^2 \cos^2\alpha \sin^2\phi~.
\end{equation}
For the top-quark, $Y_1=Y = Q-T_3$, yielding the topcolor contribution
\begin{equation}
\langle T_3 Y' \rangle_{top} = - \langle T_3 T_3\rangle_{top} \cos^2\phi =
-v^2 \sin^2\alpha \cos^2\phi~.
\end{equation}
Combining these two contributions, and simplifying, we find
\begin{equation}
\langle T_3 Y' \rangle = v^2(\cos^2\alpha -\cos^2\phi)~.
\end{equation}

Lastly, we consider the bottom right-hand element $\langle Y' Y' \rangle$.
This element receives contributions from the $U(1)_1 \times U(1)_2$ breaking
at scale $u$, as well as from electroweak symmetry breaking at scale $v$. In what
follows, we will be interested in computing the mass and couplings of the light $Z$ boson
including corrections up to order $v^2/u^2$ and the mass and couplings of the heavy
$Z'$ boson to leading order only. Thus, we need only keep the contribution
to this matrix element arising from scale $u$ -- an analysis completed in section \ref{Sec:BreakingSymmetry}
above. Comparing with eqn. (\ref{primeMassesZ}), we see that
\begin{equation}
\langle Y' Y' \rangle \approx p^2 u^2~.
\end{equation}

For the mass-squared  matrix of the $Z$ and $Z'$ bosons, we then find
\beq
{\mathsf m}_Z^2= \frac{e^2 p^2 u^2}{4 \sin^2{\theta}\cos^2{\theta}} 
\left(
\begin{array}{cc}
x^2 & x^2 \frac{\sin{\theta}}{\sin{\phi}\cos{\phi}} \left(\cos^2{\alpha} - \cos^2{\phi}\right)\\
x^2 \frac{\sin{\theta}}{\sin{\phi}\cos{\phi}} \left(\cos^2{\alpha} - \cos^2{\phi}\right) & \frac{\sin^2{\theta}}{\sin^2{\phi}\cos^2{\phi}}
\end{array}
\right),\label{ZZprimeMass}
\eeq
where we have introduced the small parameter $x^2=v^2/p^2 u^2 \ll 1$.
Note that the smallness of $x^2$ will be justified in section \ref{EWConstraints}, when we apply experimental bounds from electroweak measurements to our model. 

The mass eigenstates of the mass matrix (\ref{ZZprimeMass}) to leading order in $x^2$  are then:
\begin{align}
Z^\mu \longrightarrow Z^\mu - x^2 \frac{\sin{\phi}\cos{\phi}}{\sin{\theta}}\left(\cos^2{\alpha}-\cos^2{\phi}\right) Z'^\mu,\\
Z'^\mu \longrightarrow Z'^\mu + x^2 \frac{\sin{\phi}\cos{\phi}}{\sin{\theta}}\left(\cos^2{\alpha}-\cos^2{\phi}\right) Z^\mu.
\end{align}
From now on we will always mean the above mass eigenstates when we refer to $Z$ or $Z'$.
Expanding the masses of the $Z$- and $Z'$-bosons from the mass matrix to linear order in $x^2$
we find
\beq
M_Z^2= \frac{e^2 v^2}{4 \sin^2{\theta}\cos^2{\theta}}\left(1 - x^2 \left(\cos^2{\alpha-\cos^2{\phi}}\right)^2\right).\label{zMassCorr}
\eeq
and, in agreement with (\ref{primeMassesZ})
\beq
M_{Z'}^2 = \frac{e^2 p^2 u^2}{4 \cos^2{\theta} \sin^2{\phi}\cos^2{\phi}}=
\frac{p^2 u^2}{4} \left({g'}_1^2+{g'}_2^2\right)\,. \label{ZprimeMass2}
\eeq

By inserting the mass eigenstate of the $Z$ boson in our expression for the covariant derivative (\ref{covDerivative}), we see that the coupling of the $Z$ to fermions gets shifted to:
\begin{align}
g_Z &=\frac{e}{\sin{\theta}\cos{\theta}} \left[T^3 - Q \sin^2{\theta}-x^2 \left(Y_1-\sin^2{\phi} Y\right)\left(\cos^2{\alpha} - \cos^2{\phi}\right)\right]\label{GenZCoupling}\\
&=\frac{e}{\sin{\theta}\cos{\theta}} \left[T^3 - Q \sin^2{\theta}-x^2 Y \ \cos^2{\phi} \left(\cos^2{\alpha} - \cos^2{\phi}\right)\right],\label{gZ}
\end{align}
where in the second line we exploited the fact that $Y=Y_1$ for all fermions in our hypercharge-universal topcolor model. The coupling of the $Z'$ to fermions is
\begin{equation}
g_{Z'}=\frac{e}{\cos{\theta}\sin{\phi}\cos{\phi}}\left(Y_1-\sin^2{\phi}Y\right) =\frac{e \cos{\phi}}{\cos{\theta}\sin{\phi}}  \ Y = \sqrt{4\pi\kappa_1}\, Y
\label{gZ'}
\end{equation}
where we neglect order $x^2$ corrections since $Z'$ exchange is already suppressed relative to $Z$ exchange by the much heavier $Z'$ mass ($M_Z^2 / M_{Z'}^2 \propto x^2$).
Therefore the form of the $Z'$ effective Lagrangian remains as in eqn.  (\ref{effLag2}).

\subsection{Calculation of the Electroweak Parameters} \label{EWparCalc}

A common parametrization of universal 
corrections \footnote{As discussed in \cite{Barbieri:2004qk}, only four of the most general seven 
parameters at order $p^4$ are needed to describe universal corrections in 
models with a gap between $M_W$ and the new physics scale.} to Standard Model electroweak
scattering processes is provided by the quantities $S$, $T$, $\Delta\rho$ and $\delta$,
which are defined \cite{Chivukula:2004af} with reference to general expressions for the tree-level neutral current 
\beqa
-{\cal M}_{NC} = e^2 \frac{Q Q'}{P^2} 
& + &
\dfrac{(T^3-s^2 Q) (T'^3 - s^2 Q')}
	{\left(\frac{s^2c^2}{e^2}-\frac{S}{16\pi}\right)P^2 +
		\frac{1}{4 \sqrt{2} G_F}\left(1-\alpha T + \frac{\alpha \delta}{4 s^2 c^2}\right)
		} 
\label{NCamplitude} \\ \nonumber & \ \ & \\
&+&
\sqrt{2} G_F \, \frac{\alpha \delta}{s^2 c^2}\, T^3 T'^3 
+ 4 \sqrt{2} G_F  \left( \Delta \rho - \alpha T\right)(Q-T^3)(Q'-T'^3),
\nonumber 
\eeqa
and charged-current electroweak scattering amplitudes 
\begin{align}
  - {\cal M}_{\rm CC}
  =  \dfrac{(T^{+} T'^{-} + T^{-} T'^{+})/2}
             {\left(\dfrac{s^2}{e^2}-\dfrac{S}{16\pi}\right)P^2
             +\frac{1}{4 \sqrt{2} G_F}\left(1+\frac{\alpha \delta}{4 s^2 c^2}\right)
            }
        + \sqrt{2} G_F\, \frac{\alpha  \delta}{s^2 c^2} \, \frac{(T^{+} T'^{-} + T^{-} T'^{+})}{2},
\label{CCamplitude}
\end{align}
with $P^2$ a Euclidean momentum-squared.  In this section, we calculate how these
parameters depend on the beyond-the-standard-model degrees of freedom in our
topcolor theory: $\left[\sin{\alpha}, \sin{\phi}, u, p\right]$ or $\left[f_t, \kappa_1, u, p\right]$.   

We immediately note that
$\delta=0$ in our model, since the model does not include an extra $SU\!\left(2\right)$ triplet state.
Thus, the only corrections to charged-current scattering arise from the $S$ parameter.

The specific form of the neutral current amplitude in our model includes contributions from photon, $Z$, and $Z'$ exchange
\beq
-{\cal M}_{NC}=e^2\frac{Q Q'}{P^2} + \frac{g_Z^2}{P^2+M_Z^2} + \frac{g_{Z'}^2}{M_{Z'}^2}.
\eeq
Using the previously-derived expressions for $g_Z$ (\ref{gZ}) and $M_Z$ (\ref{zMassCorr}), we can write the $Z$-exchange term in terms of the model parameters as
\begin{equation}
\frac{g_Z^2}{P^2+M_Z^2} = 4 \mu_z^2 \left(1+\Delta_1\right)^2 \frac{\left(T^3-s^2 Q\right)\left(T'^3-s^2Q'\right)}{P^2+v^2 \mu_Z^2 \left(1-\Delta_2\right)}
\end{equation}
where we have defined
\begin{align}
\Delta_1&\equiv x^2\cos^2{\phi}\left(\cos^2{\alpha}-\cos^2{\phi}\right) \nonumber\\
\Delta_2 &\equiv x^2 \left(\casq-\cpsq\right) \\
\mu_Z^2 &\equiv  \frac{e^2}{4 \stsq \ctsq} \nonumber
\end{align}
and also derived the relationship
\beq
s^2 = \frac{\sin^2\theta + \Delta_1}{1 + \Delta_1} \,.
\eeq
Comparing the coefficient of the momentum-squared in the denominator with the equivalent expression in (\ref{NCamplitude}) yields the following expression for $\alpha S$ to leading order in $x^2$:
\beq
\alpha S = 4 x^2 \left(\casq - \cpsq\right) \cpsq \ctsq \label{eq:svall}
\eeq 
Likewise, comparing the constant term in the denominator with the equivalent expression in (\ref{NCamplitude}) tells us
\beq
\alpha T = x^2\left(\cos^4{\alpha} - \cos^4{\phi}\right) \label{eq:tvall}
\eeq

To calculate $\Delta \rho$, we consider the $Z'$ exchange contribution to the neutral currents, and recall our previously-derived expressions for the $Z'$ mass (\ref{ZprimeMass2}) and coupling constant (\ref{gZ'})
\begin{equation}
\frac{g_{Z'}^2}{M_{Z'}^2}= x^2 \frac{4}{v^2} \cos^4{\phi} \ Y Y' = \frac{4}{v^2} \left(\Delta\rho - \alpha T\right) Y Y'\,.
\end{equation}
Inserting our expression for $\alpha T$ enables us to isolate $\Delta\rho$:
\beq
\Delta\rho=x^2\cos^4{\alpha}\,.  \label{eq:rhovall}
\eeq
There is also a contribution to $\Delta\rho$ from topgluon exchange across the top and bottom quark loops of the $W$ and $Z$ vacuum polarization diagrams (cf. \cite{Popovic:1998vb})
\beq
\Delta\rho^C \approx \frac{16 \pi^2 \alpha_Y}{3 \sin^2{\theta_W}} \left(\frac{f_t^2}{M_C M_Z}\right)^2 \kappa_3.
\eeq
But due to the large topgluon mass, this contribution to $\Delta\rho$ turns out to be negligible compared to the tree-level $Z'$ contribution.  

\subsection{The Weak-Mixing-Angle}\label{WeakMixingAngle}

Next, we need to relate the theoretical quantity $\sin\theta$, which appears in our expressions for $S$, $T$, and $\Delta\rho$, to observables.   For most purposes, we will find it most convenient to use a definition that relies on the best-measured electroweak quantities, $\alpha$, $G_F$, and $M_Z$
\beq
\sin^2{\theta_Z} \cos^2{\theta_Z}=\frac{\pi \alpha}{\sqrt{2}G_F M_Z^2}.
\eeq
Recalling $v^2=1/\sqrt{2}G_F$ and using the $Z$ mass from (\ref{zMassCorr}), we find to leading order in $x^2$
\beq
\sin^2{\theta_Z}=\sin^2{\theta}+x^2 \frac{\stsq \ctsq}{\ctsq-\stsq}\left(\casq-\cpsq\right)^2 \label{thetaZ}\,.
\eeq
In studying the shift of the $W$ mass, it is easier to use the Sirlin definition of the weak mixing angle, which may be evaluated in our model using eqns. (\ref{zMassCorr}) and (\ref{wMass})
\beq
\cos^2{\theta_W} = \frac{M_W^2}{M_Z^2} = \ctsq + x^2 \ctsq \left(\casq-\cpsq\right)^2 \,. \label{thetaW}
\eeq
Combining this with eqn. (\ref{thetaZ}) yields
\beq
\cos^2{\theta_W} = \cos^2{\theta_Z} + x^2 \frac{\cos^4{\theta_Z}\left(\casq - \cpsq\right)^2}{\cos^2{\theta_Z}-\sin^2{\theta_Z}}\,, \label{eq:costheWZ}
\eeq
In deriving this last relation, we took advantage of the fact that the difference between $\sin^2{\theta}, \sin^2{\theta_Z}, s^2$ and $\sin^2{\theta_W}$ is of order $x^2$, so that any definition of the weak-mixing-angle may be used inside terms proportional to $x^2$.

\section{Precision Electroweak Fits}

\subsection{Observables and Model Parameters}\label{EWConstraints}

Although the Standard Model is only a low-energy effective theory, it nonetheless describes all data at accessible energies so well that the deviations new physics would cause in electroweak observables are constrained to be no larger than one-loop corrections within the Standard Model.  In testing the predictions of our topcolor models against the precision electroweak data, we will therefore write the theoretical values of observables as the one-loop SM value plus the tree-level corrections from the new physics:
\beq
 O^{th}=O_{SM}^{1-loop}+\delta O_{new}. \label{replacement1}
\eeq
or, equivalently,
\beq
O^{th}=O_{SM}^{1-loop}\left(1+\frac{\delta O_{new}}{O_{SM}^{tree}}\right),\label{replacement2}
\eeq
where in the denominator of the correction term, $O_{SM}^{tree}$ suffices at the required level of accuracy.  

For any specific observable $O_i$, it is conventional to parametrize the corrections due to new physics in terms of $\alpha S$, $\alpha T$, $\alpha \delta$ and $\Delta\rho$:
\beq
O_i^{th}=O_i^{1-loop}+ a_i \alpha S + b_i \alpha T + c_i \alpha \delta + d_i \Delta\rho. \label{Oth}
\eeq
where the coefficients $a_i\,,b_i\,,c_i\,,d_i$ are specific to that observable.  Since the oblique corrections $\alpha S$ and $\alpha T$ are defined as corrections to the SM with some particular reference Higgs mass, we must specify a value of $M_{H,ref}$.  While dynamical models of electroweak symmetry breaking, such as topcolor-assisted technicolor, do not include a Higgs boson as such, the Higgs doublet's role in unitarizing the scattering of longitudinal electroweak gauge bosons is played by composite resonances like the techni-rho \cite{Matsuzaki:2006wn, Sekhar Chivukula:2007ic}.  As these resonances have masses of order a TeV, we will use $M^{H,ref} = 800,\,1500$\, GeV.

Since our objective is to determine how precision electroweak data constrain our topcolor model, we use the previously defined relationships (\ref{eq:svall}, \ref{eq:tvall}, \ref{eq:rhovall}) to re-write the theoretical prediction for each observable in terms of the model's parameters.  The four extra degrees of freedom in the electroweak sector were previously denoted as $\kappa_1, u, p, f_t$; here, it is more convenient to replace $\kappa_1$ by $\cos\phi$ and $f_t$ by $\cos\alpha$.  In addition, we find that the $O^{th}_i$ depend on the product $pu$ rather than on the two separate variables; accordingly, we replace them by the previously-defined quantity  $x \equiv v / p u$.  Thus the $O^{th}_i$ depend explicitly only on three model parameters: $\cos\phi,\, \cos\alpha,$ and $x$. For example, we obtain the following expression for the $Z$ decay width to any species of charged lepton
\beq
\Gamma^{th}_{\ell^+\ell^-} = \Gamma_{\ell^+\ell^-,\,SM}^{1-loop}\left[1+x^2\left(1.1962 \cos^4{\alpha}-0.8478 \cos^2{\alpha}\cos^2{\phi}-0.3484 \cos^4{\phi}\right)\right],
\eeq
and for the total $Z$ width we have
\beq
\Gamma^{th}_{tot} = \Gamma^{1-loop}_{\rm tot,\, SM}\left[1+x^2\left(1.3506 \cos^4{\alpha}-1.5152 \cos^2{\alpha}\cos^2{\phi}+0.1646 \cos^4{\phi}\right)\right]. \label{Zdecayth}\,.
\eeq
In each case, the one-loop SM result is obtained using ZFITTER\cite{Arbuzov:2005ma}\cite{Bardin:1999yd} and SMATASY \cite{Kirsch:1994cf}, which also takes into account the phase space factors for massive final-state fermions; values are in the tables in the Appendix.
As another example, the mass shift for the $W$ boson 
\beq
M_W =   M^{1-loop}_W \left(1+ \frac{1}{2}x^2 \frac{\cos^2{\theta_Z}\left(\casq - \cpsq\right)^2}{\cos^2{\theta_Z}-\sin^2{\theta_Z}}\right). \label{mWth}
\eeq
is obtained by starting from eqn. (\ref{eq:costheWZ}).

While the $O^{th}_i$ depend on three model parameters, as discussed above, it happens that we can capture the essential physics with a two-parameter fit to $x$ and $\cos\phi$ by setting $\cos\alpha$ to a fixed value. The Pagels-Stokar formula \cite{Pagels:1979hd} that relates the size of the dynamically-generated top mass $m_t$ to the top-quark condensate $f_t$ and $m_t$
\beq
f_t^2=\frac{3 m_t^2}{8 \pi^2} \ln{\frac{u^2}{m_t^2}}
\eeq
may be rewritten as
\beq
\cos^2{\alpha} = 1- \frac{3 m_t^2}{8 \pi^2 v^2} \ln{\frac{u^2}{m_t^2}}, \label{csqaFixing}
\eeq
In this form, we see that $\cos\alpha$ varies only logarithmically with $u$; accordingly, we will fix $\cos^2\alpha \approx .91$ (i.e. $f_t \approx 75$ GeV), which corresponds to $u = 2$ TeV.  We have checked that the results obtained from the full three-parameter fit are consistent with those for the simpler two-parameter fit reported here.

\subsection{The Fit}\label{TheEWFit}

We have performed a two-parameter fit in $x^2$ and $\cos^2{\phi}$ for the full set of data shown in the Appendix.  Note that the fit is linear in $x^2$, but due to the occurrence of $\cos^4{\phi}$ in the theoretical expressions for the observables, non-linear in $\cos^2{\phi}$.  The search for the global minimum of the $\chi^2$ function in the physically allowed region of the two parameters ($x^2\geq 0$ and $0\leq \cos^2{\phi}\leq 1$) reveals that the minimum is actually on the boundary at $\cos^2{\phi}=0$ and $x^2=0.0035$ corresponding to a $\chi^2_{\rm min}=48.88$.   For $\cos^2{\phi}$ so near to zero, we can treat the fit as linear in $\cos^2{\phi}$ for purposes of statistical interpretation of the results.  Since we have $35$ observables and do a fit in two parameters, we have $\chi^2_{\rm min}/{\rm d.o.f.}= 48.88/33 =1.48$, corresponding to a 3.7\% of obtaining a $\chi^2_{min}$ of at least this size if our model is the correct description of the data.  For purposes of comparison, we note that when we fit the same data set to the SM, with a reference Higgs mass of 115 GeV, we obtain nearly the same probability: 3.5\% ($\chi^2_{\rm min}/{\rm d.o.f.}= 51.58/35 =1.47$), while a fit to the SM with a reference Higgs mass of 800 GeV yields a probability of essentially zero ($\chi^2_{\rm min}/{\rm d.o.f.}= 139.6/35 = 3.99$).

\begin{figure}[htb]
\begin{center}
\epsfig{file=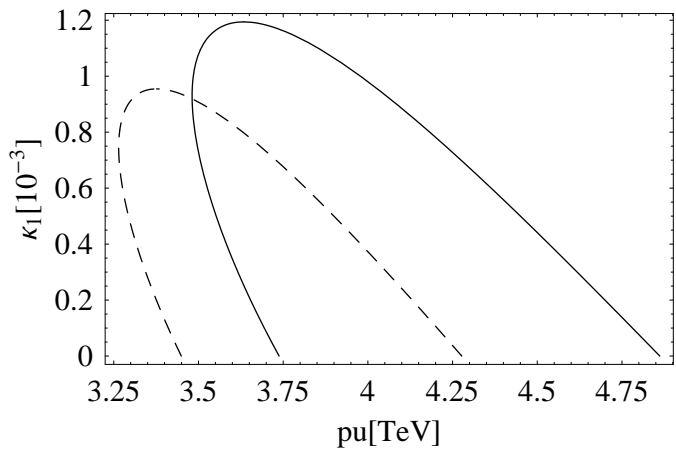,height=6cm}
\caption{\label{OwnFitsOurModel}The solid (dashed) curve shows the restriction on model parameter space obtained from a fit of all $Z$ pole and off-pole data listed in the first two tables of the Appendix for an 800 GeV (1500 GeV) reference Higgs mass. The region outside the parabola is excluded at $95 \%$ c.l.\,.}
\end{center} 
\end{figure}

To aid in interpretation, we may also rephrase our results in terms of the variables $\kappa_1$ and $pu$.  As figure \ref{OwnFitsOurModel} shows, $\kappa_1$ is restricted to a value less than $1.2\times 10^{-3}$ at 95\% c.l.\,.   It is worth recalling that the coupling of the $Z'$ boson to fermions has strength $\sqrt{4\pi\kappa_1}\, Y$ (see (\ref{gZ'})); this is of order .05 for $\kappa_1 = .001$.  At the same time, $pu$ must lie between about 3.48 TeV and 4.86 TeV.  Values of $pu$ this large are clearly in accord with our prior assumption that $x^2 \ll 1$; in fact, we find $x^2 \leq 0.005$ at 95\% c.l. \,.  Note that the presence of an upper bound on $pu$ is actually consistent with the fact that our expressions for the observables reduce to their SM values as $u \rightarrow \infty$ (rather than simply when $\kappa_1 \to 0$).    We have used a reference Higgs mass of 800 GeV in order to reflect the dynamical origin of electroweak symmetry breaking in TC2.  However, as mentioned earlier, the fit of the SM with an 800 GeV Higgs to this data is quite poor; the SM fits the current data best \cite{Alcaraz:2006mx}, for the much lower Higgs mass of approximately $80 \ {\rm GeV}$.  

\subsubsection{ Bounds on $M_{Z'}$}

\begin{figure}[htb]
\begin{center}
\epsfig{file=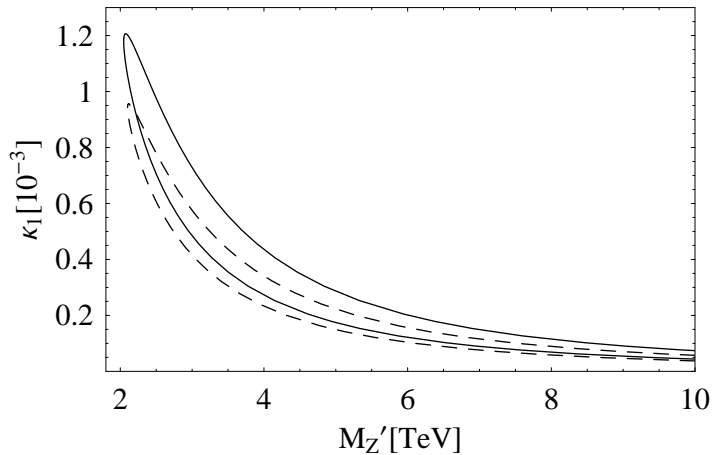,height=6cm}
\caption{\label{ZpFitsOurModel}  The solid (dashed) curve shows limits on the $Z'$ mass as a function of $\kappa_1$ resulting from a fit of all $Z$ pole and off-pole data listed in the first two tables of the Appendix for an 800 GeV (1500 GeV) reference Higgs mass.  Values outside the boomerang shaped region are excluded at $95 \%$ c.l.\,.}
\end{center} 
\end{figure}

Translating the constraint on $pu$ into a limit on the $Z'$ boson mass as a function of $\kappa_1$ using 
 eqn.\ (\ref{kappaMasses}) yields the results shown in Fig. \ref{ZpFitsOurModel}.  We observe that $Z'$ masses less than $2.08 \ {\rm TeV}$ (2.12 TeV) are excluded at a confidence level of $95 \%$ when using a Higgs mass of $800 \ {\rm GeV}$ (1500 GeV). The data do not provide an upper bound on $M_{Z'}$ because the best-fit value of $\kappa_1$ is zero and  $M_{Z'}\propto 1/\sqrt{\kappa_1}$ for $\kappa_1$-values small compared to $\alpha_Y\approx 0.01$ (from eqn.\ (\ref{kappaMasses})).

Precision electroweak data clearly provide a stronger lower bound on $M_{Z'}$ than the contact interactions discussed in section \ref{ContactInt}.  Moreover, direct searches for $Z'$ bosons also give weaker limits than the precision data.  The present CDF limit on a ``sequential" $Z'$ boson with standard-model-strength couplings to fermions is $M_{Z'} > 923$ GeV \cite{:2007sb};  the CDF limit on our $Z'$ would be even weaker since the couplings between the hypercharge-universal $Z'$ and fermions are smaller than those for a sequential $Z'$ by factors ranging from 1.5 for right-handed electrons to 15 for left-handed quarks.  We estimate $(\sigma\cdot B)_{TC2} \approx .24 (\sigma\cdot B)_{seq}$, implying that the direct lower bound on the $Z'$ mass is of order 800 GeV.  By way of comparison, we note that the limit lies between the CDF limit on the E6 $Z'$ bosons $Z'_I$ and $Z'_\chi$, as might be expected from a comparison of the various $Z'$ bosons' couplings along the lines discussed in \cite{Carena:2004xs}.

\subsubsection{Interpretation using standard electroweak parameters}

Our bounds may also be understood in terms of the standard parametrization of the electroweak corrections which we calculated in section \ref{EWparCalc}:
\bdm
\begin{split}
\alpha S &= 4 x^2 \left(\casq - \cpsq\right) \cpsq \ctsq\\
\alpha T &= x^2\left(\cos^4{\alpha} - \cos^4{\phi}\right)\\
\Delta\rho&=x^2\cos^4{\alpha}.
\end{split}
\edm
Since the best fit value for $\cos^2{\phi}$ is zero, $\Delta\rho\approx\alpha T$ and $S\approx 0$. Fixing $\cos^2{\alpha}$ by the Pagel-Stokar relation, eqn.\ (\ref{csqaFixing}), gives $T\approx 0.4$ for the best fit value $x^2=0.0035$.   
The experimental bounds on $S$ and $T$ are commonly illustrated as an ellipse in the $S$-$T$ plane for a given reference Higgs mass. Some of these plots (See, e.g., Fig.\ E.2 in ref.\ \cite{Alcaraz:2006mx}.) also show how contributions from a very large Higgs mass would pull the Standard Model prediction outside the region defined by the experimental bounds. For a (reference) Higgs mass of around $1 \ {\rm TeV}$ new physics adding $\sim 0.4$ to the value of $T$ is required to move the theoretical prediction back into the experimentally-favored ellipse.  This is why the presence, in our model, of new physics that persists in the limit $\cos^2{\phi} \to 0$ is important.

\subsubsection{The gap triangle and $M_c$}

Generating the desired top-quark condensate (and only a top condensate) has placed the model within the gap triangle in the $\kappa_3$-$\kappa_1$ plane (Fig.\ \ref{GapTriangleGauged}).  Moreover, as discussed in section \ref{GapEquations}, the known top mass confines $\kappa_1$ and $\kappa_3$ to the region near the left-hand side of the gap triangle (which includes the lower tip) when the scale $pu$ is of order a TeV or more.   We have now found that $pu \gtrsim 3 \ {\rm TeV}$ and also that $\kappa_1\lesssim 10^{-3}$. This means the model is now restricted to the lower tip of the gap triangle, where $\kappa_3 \approx 1.9$ is close to its critical value for the gauged NJL approximation to the gap equations.

We now consider the implications for the mass of the topgluon.  Examining eqn.\ (\ref{kappaMasses}) we see that $M_c$ depends linearly on $u$ but  less strongly on $\kappa_3$. Moreover, the discussion above has made clear that the $\kappa_3$ must have a value very close to 1.9.  Fixing $\kappa_3$ and allowing $pu$ to vary in the range suggested by Fig. \ref{OwnFitsOurModel} (3.48 TeV $\leq pu \leq$ 4.86 TeV), we estimate 18 TeV $\leq M_c \leq$ 25 TeV for $p=1$.  This is well above the bound we obtained from FCNC in section \ref{sec:FCNC}.

\subsubsection{Effects of top-pions in TC2 models}
\label{sss:effect}

As discussed in \cite{Burdman:1997pf}, top-pion exchange affects the value of $R_b$ predicted by TC2 models, with the shift depending markedly on the mass assumed for the top-pions.   We have performed an additional fit to the Hypercharge-Universal TC2 model, including this effect for various top-pion masses.  

For a given top-pion mass, we read (or extrapolated) the fractional shift in $R_b$ from the first figure in \cite{Burdman:1997pf}, and modified the predicted value of $R_b^{th}$ accordingly.  Because the shift in $R_b$ tends to decrease the predicted value relative to the SM prediction, the fit probability for the minimum value of $\chi^2$ is reduced.  For example, assuming $M_{\pi_t} \approx 900$ GeV, the fit probability becomes 0.25\%; the top-pion mass for which the fit probability is 1\% is $M_{\pi_t} \approx 1200$ GeV.   As shown in figure 5, the allowed range of $\kappa_1$ and $M_{Z'}$ is very slightly altered  by inclusion of this shift in $R_b$, such that the lower bound on $M_{Z'}$ becomes 2.05 TeV. 

However, as also discussed in  \cite{Burdman:1997pf}, while the presence and sign of the effect of top-pions on $R_b$ is certain,  the size of the effect is not.  In fact, it is quite sensitive to several factors that are hard to evaluate precisely: the value of $f_{t}$, the value of $m_{ETC}$, and a radiative correction factor due to topcolor gauge interactions.   For instance, as \cite{Burdman:1997pf} notes, if the value of $f_t$ were about twice as large as the Pagels-Stokar formula suggests, then the shift in $R_b$ would be smaller by a factor of 6.7.  In this case, Hypercharge-Universal TC2 would yield a fit probability of 1\% for $M_{\pi_t} \approx 250$ GeV and 3\% for $M_{\pi_t} \approx 900$ GeV.  As shown in figure 5, the allowed  region of $\kappa_1$ and $M_{Z'}$ is noticably shifted by the changed value of $f_t$ (but still not much affected by the value of $M_{\pi_t}$), such that the lower bound on $M_{Z'}$ becomes 1.68 TeV. 

Overall, we conclude that the Hypercharge-Universal TC2 model is consistent with the data at hand, within the theoretical uncertainties discussed above.   As indicated by Figure 5, the bounds on $\kappa_1$ and $M_{Z'}$ are not terribly sensitive to these theoretical uncertainties, and the $Z'$ should have a mass and coupling that render it accessible to LHC experiments.

\begin{figure}[htb]
\begin{center}
\epsfig{file=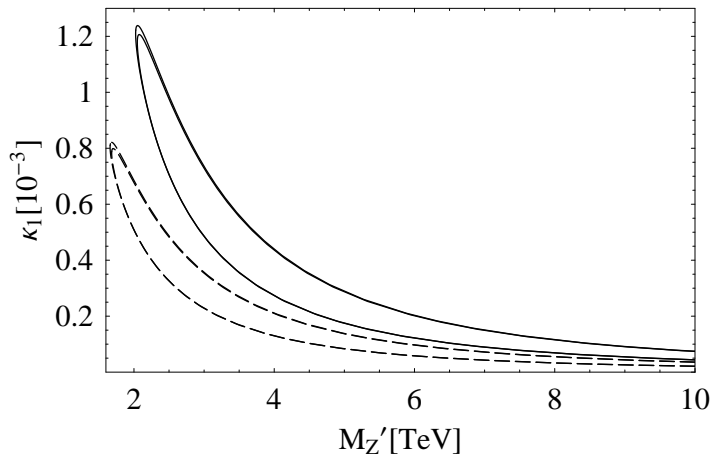,height=6cm}
\caption{\label{ZpFitsOurModelRb}  Limits on the $Z'$ mass as a function of $\kappa_1$ resulting from a fit of all $Z$ pole and off-pole data listed in the first two tables of the Appendix for an 800 GeV reference Higgs mass.  Values outside the appropriate boomerang shaped region are excluded at $95 \%$ c.l.\,.  The overlapping pair of solid curves corresponds to $f_t = 75$ GeV;  the upper one includes exchange of a top-pion of mass 1200 GeV and lower one does not include top-pion exchange. The overlapping pair of dashed curves corresponds to $f_t = 150$ GeV; the upper one includes exchange of a top-pion of mass 250 GeV and the lower one does not include top-pion exchange.  Note that the main effect on the shape and position of the allowed region comes from altering $f_t$.}
\end{center} 
\end{figure}

\section{Limits on the Classic and Flavor-Universal Topcolor Models}\label{SimmonsModel}

For comparison, we briefly look at the Classic and Flavor-Universal TC2 models, which include flavor non-universal $Z'$ bosons.  Electroweak precision limits on these models were previously obtained in ref. \cite{Chivukula:2002ry}; at that time, there was still a narrow window of parameter space in which a $Z'$ mass below 1 TeV was possible.

Starting from the general expression for the $Z$ coupling to fermions in eqn.\ (\ref{GenZCoupling}) and using the fermion charge assignments of Table \ref{assignments-old}, we see that in the Classic and Flavor-Universal TC2 models,  the corrections to the Standard Model expressions for  third generation fermions are proportional to $\cos^2{\phi}$, while the first and second generation particles receive a correction proportional to $\sin^2{\phi}$.  Knowing this enabled us to fit the models with flavor non-universal $Z'$ bosons to the data in the last two tables in the Appendix;  note that we must now employ the data that does not, a priori, assume generation universality.
  
Searching for the global minimum of $\chi^2$ in the physical region where $x^2\geq 0$ and $0\leq \cos^2{\phi}\leq 1$ (and setting $f_t\approx75 \ {\rm GeV}$), we find $\chi^2_{\rm min}/{\rm d.o.f.}= 106/39= 2.71$, corresponding to a probability of order $10^{-8}$.   The best-fit value of $\cos^2\phi$ is close to 0, which forces $\sin^2\phi$ to be of order 1, thereby increasing the difference between the $Z$ boson's couplings to fermions in the third generation and those in the first or second generations; given the degree of lepton universality displayed by the $Z$-pole data, it is not surprising that a poor fit results.  Including the shift towards more negative values of $R_b$ due to top-pion exchange decreases the fit probability still further. The flavor non-universal Z' bosons of these TC2 models are simply not concordant with the precision electroweak data.

\section{Conclusions}\label{Conclusions}

We have introduced a new topcolor-assisted technicolor with a flavor-universal $Z'$ boson, shown that it is capable of producing a top-quark condensate, and studied its phenomenology.  Our analysis shows that precision electroweak measurements (including $R_b$) provide tight constraints on this model.  In the narrow region of  parameter space where these constraints are satisfied, the bounds from FCNC, contact interactions, and $U(1)$ triviality are automatically obeyed.   In contrast, we find that precision electroweak constraints now exclude models with flavor-non-universal $Z'$ bosons \cite{Hill:1994hp}\cite{Popovic:1998vb}.

Our TC2 model with a flavor universal hypercharge sector can fit the electroweak data as well as the SM with a light Higgs boson.  However, the topcolor-symmetry-breaking scale is driven to the value $pu \approx 4 \ {\rm TeV}$, which means that $m_t$ is much lighter than the scale of the dynamics by which it is generated and implies a need for fine tuning of the coupling factors $\kappa_1$ and $\kappa_3$.  The topgluon coupling factor $\kappa_3$ is constrained to lie near its critical value of 1.9; accordingly, 
the topgluon mass is of order 20 TeV: too heavy for existing or planned colliders to access directly. 
The $Z'$ coupling factor $\kappa_1$ is of order $.001$ or smaller (the $Z'$ couples to fermions as $\sqrt{4\pi\kappa_1}\,Y$).  The $Z'$ mass must therefore lie above 1.6 - 2 TeV: too heavy for direct production at the Tevatron, but within reach of LHC.

\section{Acknowledgements}

The authors thank M. Gr\"unewald and S. Clark for useful conversations.
This work was supported in part by the US National Science Foundation under
grant  PHY-0354226.  F.B. and M.F. acknowledge support from the German National Academic Foundation (Studienstiftung des deutschen Volkes).  R.S.C. and E.H.S. acknowledge the 
Galileo Galilei Institute for support during the completion of this work.

\appendix*
\section{Experimental Data and SM Predictions Used In Our Analysis}

This appendix shows the observables used to assess each model, along with the experimentally measured values and the predicted values we obtained.  The first pair of tables shows Z-pole and LEP II observables as calculated in the hypercharge-universal model introduced in this paper.  The second pair contrasts our fit to the SM with a light Higgs boson; the third pair contrasts the results for the models with  non-flavor-universal $Z'$ bosons.

\begin{figure}[tb]
\begin{center}
\epsfig{file=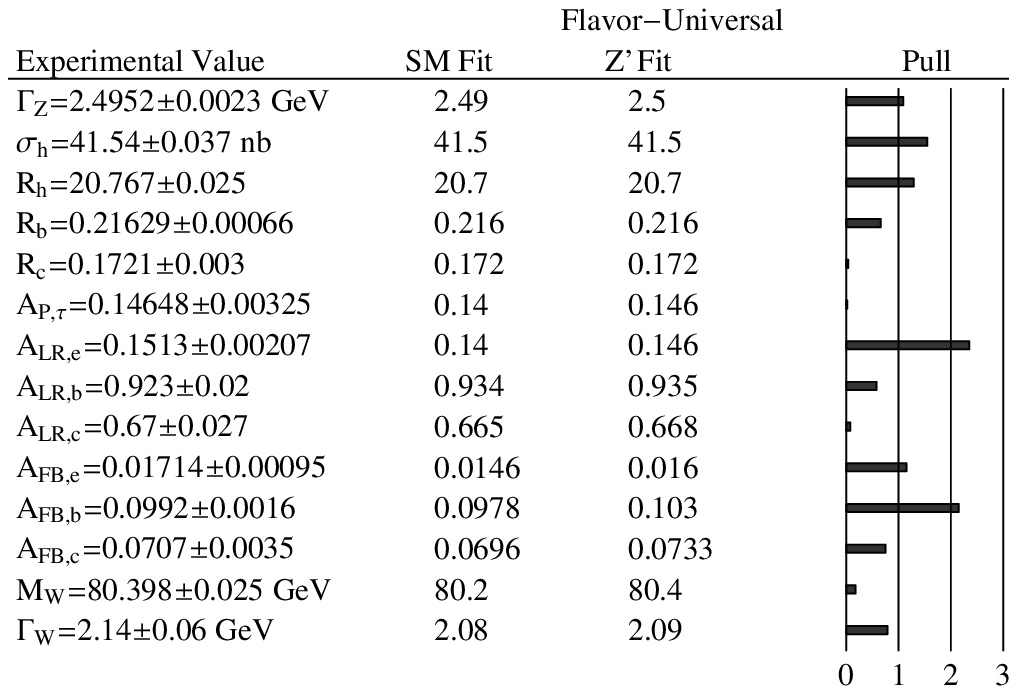,width=10cm}
\caption{\label{fig:hypuniv-Z}   Fits to Z-pole observables in the Hypercharge-Universal TC2 model introduced in this paper.  From left to right, the columns show the experimental values, the 1-loop SM values ${\cal{O}}^{1-loop}_i$ with $M_{H,ref} = 800$ GeV, and the predictions for the TC2 model with their pulls.  The TC2 model fit assumed $f_t = 75$ GeV and $M_{H,ref} = 800$ GeV.}
\end{center} 
\end{figure}

\begin{figure}[bt]
\begin{center}
\epsfig{file=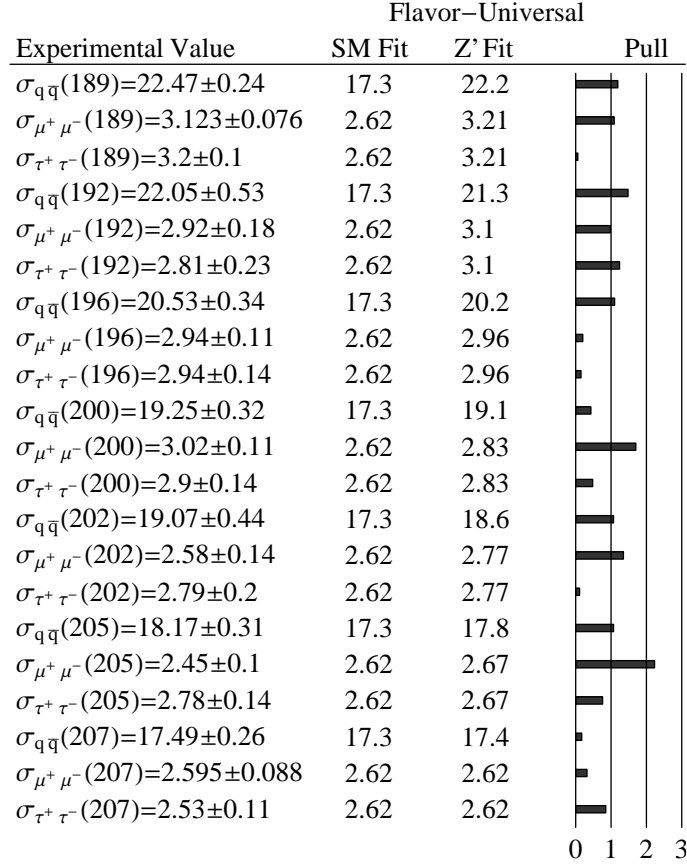,width=9cm}
\caption{\label{fig:hypuniv-LEPII}  Fits to LEP II observables at several values of $\sqrt{s}$ in the Hypercharge-Universal TC2 model introduced in this paper. From left to right, the columns show the experimental values, the 1-loop SM values ${\cal{O}}^{1-loop}_i$ with $M_{H,ref} = 800$ GeV, and the predictions for the TC2 model with their pulls.  The TC2 model fit assumed $f_t = 75$ GeV and $M_{H,ref} = 800$ GeV.}
\end{center} 
\end{figure}

\newpage

\begin{figure}[tb]
\begin{center}
\epsfig{file=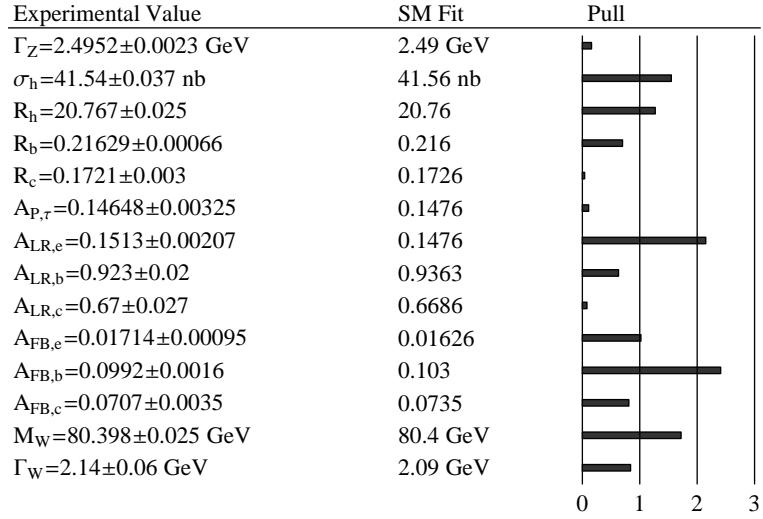,width=10cm}
\caption{\label{fig:SM-Z}   1-loop fits to Z-pole observables in the Standard Model with $M_{H,ref} = 115$ GeV, showing the experimental values, the predicted values, and the pulls.}
\end{center} 
\end{figure}

\begin{figure}[bt]
\begin{center}
\epsfig{file=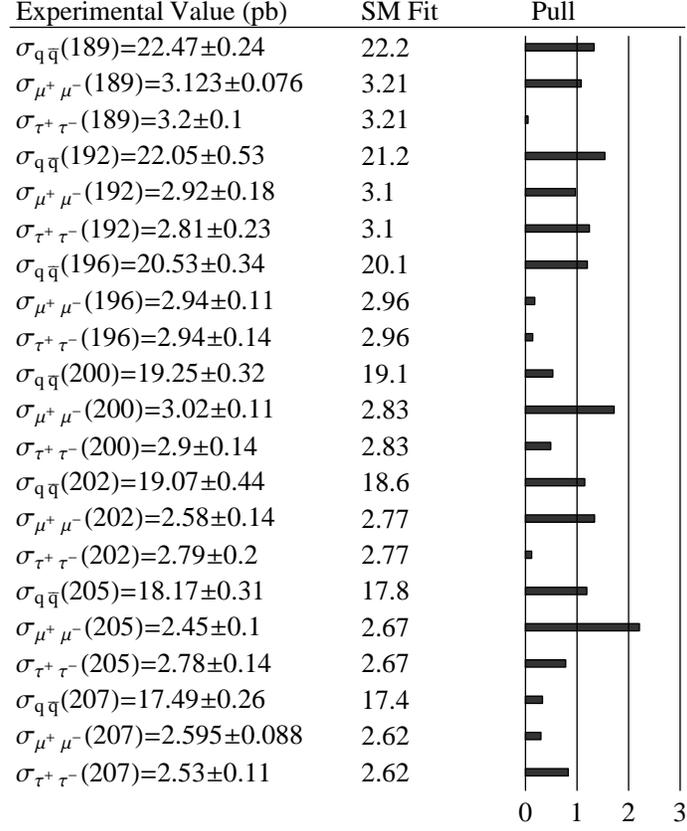,width=9cm}
\caption{\label{fig:SM-LEPII}  1-loop fits to LEP II observables at several values of $\sqrt{s}$ in the Standard Model with $M_{H,ref} = 115$ GeV, showing the experimental values, the predicted values, and the pulls.}
\end{center} 
\end{figure}

 \newpage

\begin{figure}[tb]
\begin{center}
\epsfig{file=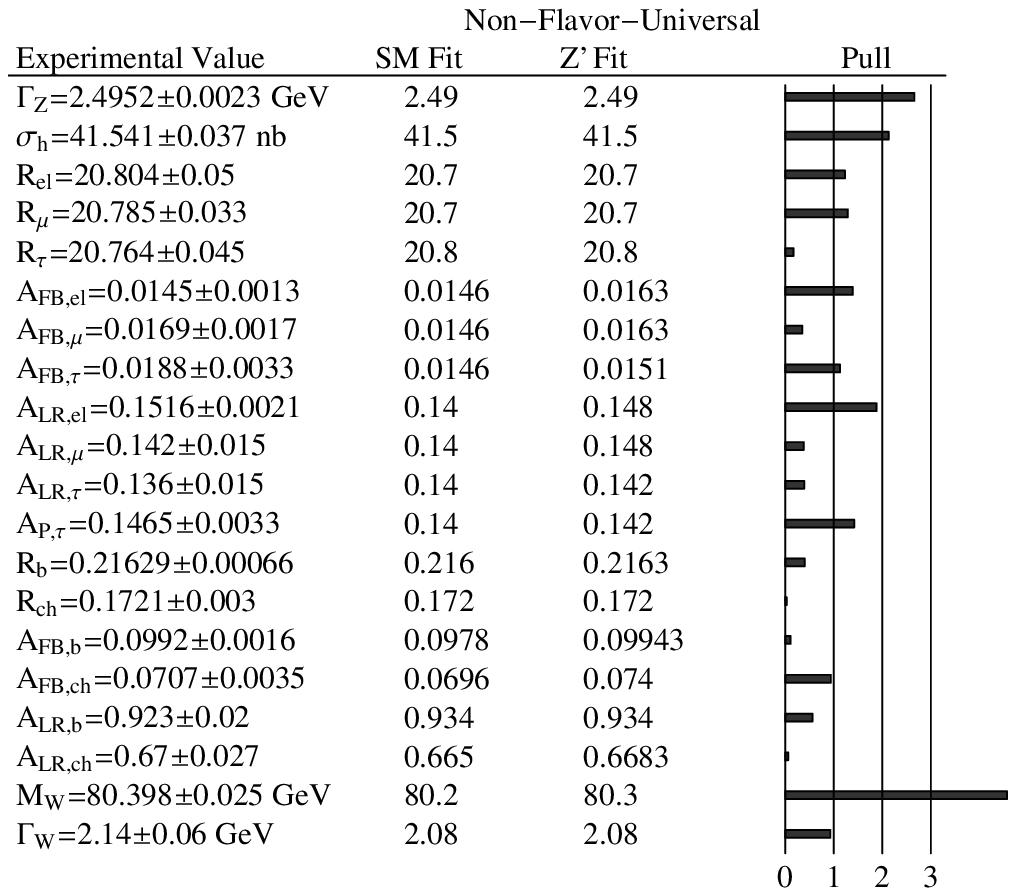,width=9.5cm}
\caption{\label{fig:hypuniv2-Z}   Fits to Z-pole observables in the Classic \protect\cite{Hill:1994hp} or Flavor-Universal TC2 \protect\cite{Lane:1998qi,Popovic:1998vb} model.  From left to right, the columns show the experimental values, the 1-loop SM values ${\cal{O}}^{1-loop}_i$ with $M_{H,ref} = 800$ GeV, and the predictions for the TC2 model with their pulls.  The TC2 model fit assumed $f_t = 75$ GeV and $M_{H,ref} = 800$ GeV.}
\end{center} 
\end{figure}

\begin{figure}[bt]
\begin{center}
\epsfig{file=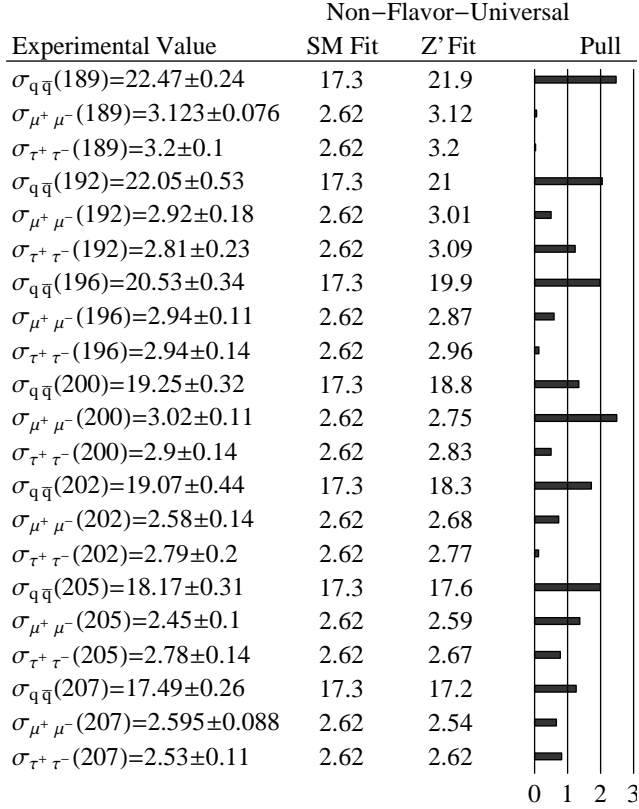,width=8.4cm}
\caption{\label{fig:hypuniv2-LEPII}  Fits to LEP II observables at several values of $\sqrt{s}$  in the Classic \protect\cite{Hill:1994hp} or Flavor-Universal TC2 \protect\cite{Lane:1998qi,Popovic:1998vb} model.  From left to right, the columns show the experimental values, the 1-loop SM values ${\cal{O}}^{1-loop}_i$ with $M_{H,ref} = 800$ GeV, and the predictions for the TC2 model with their pulls.  The TC2 model fit assumed $f_t = 75$ GeV and $M_{H,ref} = 800$ GeV.}
\end{center} 
\end{figure}


\end{document}